\documentclass[twocolumn,superscriptaddress]{revtex4}
\usepackage{graphicx}

\begin{document}
\title{Highly efficient second-harmonic generation from indefinite epsilon-near-zero slabs of subwavelength thickness}

\author{A. Ciattoni}
\affiliation{Consiglio Nazionale delle Ricerche, CNR-SPIN 67100 L'Aquila}

\begin{abstract}
We theoretically predict efficient optical second-harmonic generation (SHG) from a few hundred nanometer thick slab consisting of a quadratic
nonlinear anisotropic medium whose linear principal permittivities are, at the fundamental wavelength, very small and have different signs
(indefinite medium). We show that, by illuminating the slab with a p-polarized fundamental wave (with intensity of a few $\rm{MW/cm^2}$), a highly
efficient scattering of the second-harmonic field occurs when the conditions of linear complete slab transparency for the fundamental wave are met.
The high efficiency of the SHG process, stems from the large non-plasmonic enhancement of the longitudinal field, perpendicular to the slab surface,
produced by the very small value of the slab dielectric permittivities. A suitable nano-structured composite is proposed and numerically designed for
observing the novel non-phase-matched and highly efficient SHG process from nano-structures.
\end{abstract}

 \maketitle

Enhancing optical second-harmonic generation (SHG) \cite{KangKa,Samant} is at present one of the most relevant task of nonlinear optics due to the
major role played by frequency-doubling in coherent green and blue light sources design \cite{Yamada}, chemistry \cite{CornCo}, biosensing
\cite{Campag}, etc. In situations where standard phase-matching or quasi-phase-matching techniques cannot be used, efficient SHG is generally
achieved by resorting to specific configurations providing a strong field enhancement such as, for example, resonant microcavities
\cite{Pelleg,CaoCao,LeiLei} or photonic crystals \cite{Mondia,Liscid,Siltan,ZhaoZa}. Analogously, the field enhancement occurring on a rough metal
surface due to the excitation of surface plasmon polaritons is responsible for a substantial enhancement of the surface SHG \cite{ChenCh}. Conceiving
subwavelength coherent light sources is a fundamental issue of modern nanophotonics \cite{Nakaya,HsiehH} so that achieving SHG from nanostructures is
an important target. At the nanometer length scale, the small interaction distances generally entail an highly inefficient SHG so that peculiar
mechanisms for locally enhancing the electromagnetic field of the fundamental wave have to be harnessed. As an example, the plasmonic field
enhancement \cite{Barnes} occurring in vicinity of the subwavelength apertures in a metallic film (accompanying the extraordinary linear optical
transmission \cite{Ebbese}) is responsible for an efficient SHG \cite{FanFan,Lesuff,KangKK,SchonS} when the holes are filled by a quadratic nonlinear
material. Another strong plasmonic field enhancement occurs within spherical nanocavities with dielectric core and plasmonic nanoshell \cite{Oldenb}
so that, if the core is filled with a noncentrosymmetric nonlinear medium, a lager enhancement of the SHG is observed \cite{PuPuPu}.

In this Letter we theoretically show that a subwavelength thick slab consisting of a quadratic nonlinear medium whose linear permittivities are very
small and have different signs is able to provide highly efficient SHG. Properties and applications of epsilon-near-zero materials have recently been
proposed both in the linear \cite{Silver,Liuuuu,Edward,Alekse,Nguyen} and in the nonlinear regime \cite{Powell,Ciatt1,Ciatt2,Ciatt3,Ciatt4} and
indefinite materials \cite{SmithS,LiuLi1,Nogino,LiuLi2} have attracted a good deal of attention as well. Here we point out that, the small value of
the dielectric permittivities produces a strong non-plasmonic enhancement of the longitudinal field perpendicular to the slab surface. The field
enhancement is simply a consequence of the fact that, in conditions of linear complete slab transparency, the longitudinal fundamental field within
the slab across the vacuum-slab interface coincides with the longitudinal field of the incident fundamental wave divided by the very small dielectric
permittivity. We predict that the enhanced longitudinal field is responsible for a SHG process with efficiency up to $40 \%$, for impinging optical
intensities of a few of $\rm{MW/cm^2}$.

\begin{figure}
\includegraphics[width=0.45\textwidth]{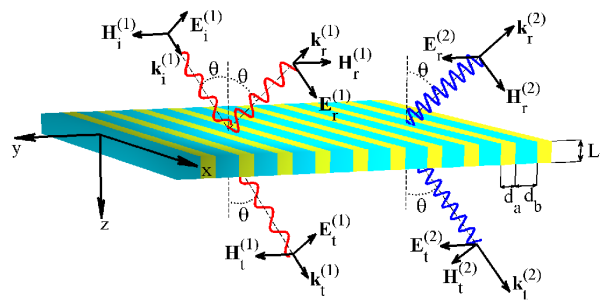}
\caption{(Color online) Slab and fields geometry for the SHG process.}
\end{figure}
The geometry of the considered SHG setup is sketched in Fig.1 and, labelling hereafter the fundamental and second-harmonic quantities with the
superscripts $(1)$ and $(2)$, we choose $\lambda^{(1)} = 827 \: \rm{nm}$ and $\lambda^{(2)} = \lambda^{(1)}/2 = 413.5 \: \rm{nm}$ for the two
wavelengths. The slab of thickness $L$ consists of quadratic nonlinear anisotropic medium whose relative dielectric tensor is $\epsilon =
\textrm{diag} \left(\epsilon_x, \epsilon_y, \epsilon_z \right)$ (i.e. its principal axes coincide with cartesian axes) and whose second-order
nonlinear optical tensor has the only non-vanishing components $d_{14} = d_{25} = d_{36} = \chi /2$. Here we choose $\epsilon_x^{(1)}=-0.01$,
$\epsilon_z^{(1)}=0.001$ and $\epsilon_y^{(2)}=2.86$ for the only permittivities playing a role in our discussion and $\chi = 1.3 \: \rm{pm/V}$ for
the nonlinear susceptibility (see below for a possible composite medium exhibiting these properties). The slab is illuminated by a plane fundamental
wave (FW) with incidence angle $\theta$ and with its electric field polarized in the plane of incidence (p-polarization) so that the reflected (r)
and transmitted (t) plane FWs are p-polarized as well. The slab also scatters reflected (r) and a transmitted (t) plane second harmonic waves (SHW)
whose electric fields are polarized perpendicular to the plane of incidence (s-polarization), as a consequence of the chosen anisotropic quadratic
nonlinearity (see Fig.1 for the definitions of the field amplitudes and wave vectors).
\begin{figure}
\includegraphics[width=0.45\textwidth]{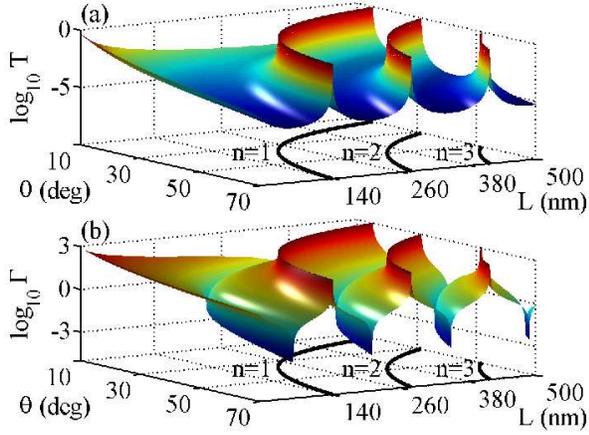}
\caption{(Color online) Logarithmic plot of the linear slab transmissivity T (panel a) and of the linear longitudinal field enhancement factor
$\Gamma$ (panel (b)) for various incidence angles $\theta$ and slab thicknesses $L$. The solid lines along the bottom planes and marked with
$n=1,2,3$ are the curves of linear complete slab transparency (i.e. $K^{(1)}L=n\pi$).}
\end{figure}

As a prelude to the analysis of the SHG process, it is worth discussing the linear slab behavior observable when the optical intensity of the
incident FW is very small. The linear slab transmissivity is
\begin{equation} \label{T}
T =\frac{|{\bf E}_t^{(1)}|^2}{|{\bf E}_i^{(1)}|^2} = \frac{1}{\displaystyle \left| \cos(K^{(1)}L) - i F \sin (K^{(1)}L) \right|^2}
\end{equation}
where $K^{(1)}(\theta) = k^{(1)} \sqrt{\epsilon_x^{(1)} (1-\sin^2\theta / \epsilon_z^{(1)} ) }$ is the wavevector of the radiation excited within the
slab, $k^{(1)} = 2\pi /\lambda^{(1)}$ and $F = k^{(1)} \epsilon_x^{(1)} \cos \theta/(2K^{(1)}) + K^{(1)} / (2k^{(1)} \epsilon_x^{(1)} \cos \theta) $.
In Fig.2(a) we plot the logarithm of $T$ as a function of both the incidence angle $\theta$ and the slab thickness $L$ from which we note that the
slab is practically always opaque to the considered radiation except for very specific incidence angles (dependent on the slab thickness) for which
it is completely transparent. From Eq.(\ref{T}) it is evident that total transmission $T=1$ occurs if $K^{(1)} L = n \pi$ and the ensuing curves of
complete transparency are reported with solid lines along the bottom plane of Fig.2(a). Note that if $K^{(1)}L$ is slightly different from $n\pi$,
the denominator in Eq.(\ref{T}) is dominated by the term containing F (since $|F| \approx |\epsilon_x^{(1)}|^{-1} \gg 1$) so that the overall
transmissivity turns out to be proportional to $|\epsilon_x^{(1)}|$ thus explaining the sharpness of the ridges of T of Fig.2(a) corresponding to
complete transparency. In Fig.2(b) we plot the logarithm of the longitudinal field enhancement factor $\Gamma = |\hat{\bf e}_z \cdot {\bf E}^{(1)}
(0^+)|/|\hat{\bf e}_z \cdot {\bf E}_i^{(1)}|$ defined as the ratio between the modulus of the field perpendicular to the slab within the medium just
across the interface $z=0$ (see Fig.1) and the modulus of longitudinal field of the {\it sole} incident FW. Note that, in correspondence of the
curves of complete slab transparency, the field enhancement factor is $\Gamma = 1000=1/\epsilon_z^{(1)}$. This is easily understood from one of the
field matching condition at the interface $z=0$ (continuity of normal component of the electric displacement field, i.e. $\hat{\bf e}_z \cdot ({\bf
E}_i^{(1)} + {\bf E}_r^{(1)} )= \epsilon_z^{(1)} [ \hat{\bf e}_z \cdot {\bf E}^{(1)} (0^+) ]$) since, when $T=1$ no reflected wave is generated (i.e.
${\bf E}_r^{(1)}=0$), so that $\hat{\bf e}_z \cdot {\bf E}^{(1)} (0^+) = [ \hat{\bf e}_z \cdot {\bf E}_i^{(1)} ] / \epsilon_z^{(1)}$. We conclude
that the considered slab provides a powerful mechanism for attaining a strong enhancement (amounting to $1/\epsilon_z^{(1)}$) of the longitudinal
field. Note that, for the field enhancement to occur it is essential that $K^{(1)}$ is real and this motivates our choice of the permittivities'
signs.

In order to discuss the SHG process we note that the incident p-polarized FW (see Fig.1) produces, within the slab, the transverse magnetic (TM)
fundamental field ${\bf E}^{(1)} = e^{i k^{(1)} x \sin \theta } [ E_x^{(1)}(z) \hat{\bf e}_x  + E_z^{(1)}(z) \hat{\bf e}_z ]$, ${\bf H}^{(1)} = e^{i
k^{(1)} x \sin \theta } [ H_y^{(1)}(z) \hat{\bf e}_y ]$ which, due to the chosen quadratic nonlinearity, couples to the transverse electric (TE)
second-harmonic field ${\bf E}^{(2)} = e^{i 2k^{(1)} x \sin \theta } [ E_y^{(2)}(z) \hat{\bf e}_y ]$, ${\bf H}^{(2)} = e^{i 2k^{(1)} x \sin \theta }
[ H_x^{(2)}(z) \hat{\bf e}_x + H_z^{(2)}(z) \hat{\bf e}_z ]$ (here we have assumed $e^{-i\omega t}$ and $e^{-i2\omega t}$ as time factors for the FW
and the SHW fields, respectively). The second-harmonic TE field within the slab matches with the s-polarized reflected and transmitted plane SHWs
scattered by the slab \cite{Higher}. The dielectric slab response is described, in the considered situation, by the electric displacement fields
\begin{eqnarray} \label{response1}
D^{(1)}_x &=& \epsilon_0 e^{i k^{(1)} x \sin \theta } \left( \epsilon^{(1)}_x E_x^{(1)}+ \chi E_z^{(1)*} E_y^{(2)} \right), \nonumber \\
D^{(1)}_z &=& \epsilon_0 e^{i k^{(1)} x \sin \theta } \left( \epsilon^{(1)}_z E_z^{(1)}+ \chi E_x^{(1)*} E_y^{(2)} \right),
\end{eqnarray}
for the FW and
\begin{equation} \label{response2}
 D^{(2)}_y = \epsilon_0 e^{i 2 k^{(1)} x \sin \theta } \left( \epsilon^{(2)}_y E_y^{(2)}+ \chi E_x^{(1)} E_z^{(1)} \right),
\end{equation}
for the SHW. After setting $I^{(1)}_i \equiv 1/2 \sqrt{\epsilon_0 / \mu_0} |{\bf E}_i^{(1)}|^2 = 7.9 \: \rm{MW/cm^2}$ for the optical intensity of
the incident FW, we have numerically solved, for various incidence angles $\theta$ and slab thicknesses $L$, the coupled sets of Maxwell equations
(within the slab $0<z<L$, where Eqs.(\ref{response1}) and (\ref{response2}) hold) for the FW and the SHW, along with matching conditions (at $z=0$
and $z=L$) describing the slab illumination by the sole incident FW. Consequently, we have evaluated the slab efficiencies $\eta_r= |{\bf
E}_r^{(2)}|^2 /|{\bf E}_i^{(1)}|^2$ and $\eta_t = |{\bf E}_t^{(2)}|^2 / |{\bf E}_i^{(1)}|^2$ of converting the incident FW into reflected (for $z<0$)
and transmitted (for $z>L$) SHWs and their logarithm plots are reported in Fig.3(a) and 3(b), respectively.
\begin{figure}
\includegraphics[width=0.45\textwidth]{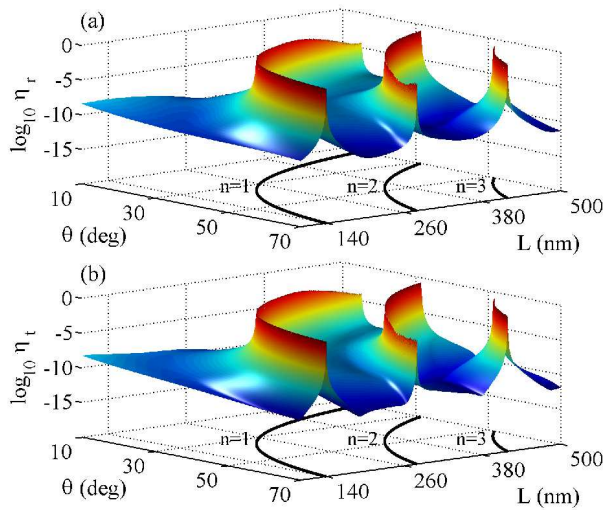}
\caption{(Color online). Logarithmic plots of the slab efficiencies $\eta_r$ (panel (a)) and $\eta_t$ (panel (b)) of converting the incident FW into
reflected and transmitted SHWs, respectively. The SHG efficiencies are evaluated for the optical intensity $I^{(1)}_i =7.9 \: \rm{MW/cm^2}$ of the
incident FW. The solid lines along the bottom planes and marked with $n=1,2,3$ are the curves of linear slab complete transparency (i.e.
$K^{(1)}L=n\pi$).}
\end{figure}
From Fig.3, it is particularly evident that, for the chosen optical intensity of the incident FW, SHG uniquely occurs for specific incident angles
and slab thicknesses practically coinciding with those providing linear total transmission of the FW (see the sharp ridges of $\eta_t$ and $\eta_r$
approximately located upon the curves $K^{(1)}L=n\pi$). The physical origin of such phenomenology is easily grasped by noting that SHG is here due to
the source term proportional to the components $E_x^{(1)}$ and $E_z^{(1)}$ of Eq.(\ref{response2}), so that, due to the small value of the nonlinear
susceptibility $\chi$, the SHW field scattered by the slab is generally very weak for the chosen impinging optical intensity $I^{(1)}_i$. In these
conditions, the nonlinear terms in Eqs.(\ref{response1}) can be generally neglected so that the FW does not experience the effect of the quadratic
nonlinearity and it consequently exhibits the linear behavior of Fig.2.
\begin{figure}
\includegraphics[width=0.45\textwidth]{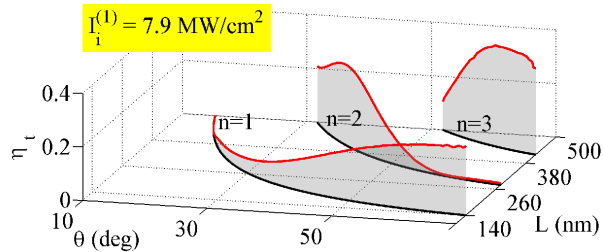}
\caption{(Color online) Values of the SHG efficiency $\eta_t$ (extracted from the numerical simulation of Fig.3) for incident angles and slab
thicknesses corresponding to the linear slab complete transparency for the FW.}
\end{figure}
Such an unattractive scenario is fundamentally altered when the conditions of linear complete slab transparency for the FW are met, since the above
described enhancement mechanism of the longitudinal field $E_z^{(1)}$ entails an enhancement of the source term $\chi E_x^{(1)} E_z^{(1)}$ so marked
to yield efficient SHG (see the sharp ridges of Figs.3(a) and 3(b)). In Fig.4, we plot the values of the above evaluated SHG efficiency $\eta_t$
exclusively along the curves of linear total transmission of the FW. From Fig.4 we conclude that the proposed mechanism is very efficient, even at
the chosen optical intensity $I^{(1)}_i = 7.9 \: \rm{MW/cm^2}$ since it is characterized by values of the efficiency $\eta_t$ up to $40 \%$ and it
takes place in slabs with thickness spanning the subwavelength range $0<L<500 \: \rm{nm}$. For the sake of completeness, note that the proposed SHG
scheme does not rely on the phase-matching condition since the slab SHW wavevector is $K^{(2)} = 2 k^{(1)} \sqrt{\epsilon_y^{(2)}-\sin^2 \theta}$ (in
the undepleted pump regime) so that the condition $2K^{(1)}=K^{(2)}$ yields the phase matching angle $\theta_{PM} = \rm{asin} \sqrt{\epsilon_z^{(1)}
( \epsilon_x^{(1)} - \epsilon_y^{(2)} ) / ( \epsilon_x^{(1)} - \epsilon_z^{(1)} )} \simeq 30.35 ^\circ$, an angle not playing any special role in the
above described process (see Figs.3 and 4).

The above analysis has been carried out for a specific value of $I^{(1)}_i$. In order to discuss the role played by the FW intensity on the
considered SHG process, we have performed further numerical simulations in a $L = 400 \: \rm{nm}$ thick slab to evaluate the SHG efficiency as a
function of the incidence angle and the optical intensity $I^{(1)}_i$. The logarithm of the obtained $\eta_t$ is plotted in Fig.5(a) (where the
$I^{(1)}_i$ axis is logarithmic) and we have considered a restricted angular range ($37^\circ < \theta < 45^\circ$) surrounding the angle $\theta =
40.9^\circ$ which is one the angles (the one belonging to the $n=2$ branch) at which the $400 \: \rm{nm}$ thick slab is linearly transparent to the
FW (see Fig.2). Note that, for angles not close to $\theta = 40.9^\circ$, the efficiency $\eta_t$ is a growing function of $I^{(1)}_i$ whereas, at
$\theta = 40.9^\circ$, the efficiency grows for intensities up to $50 \: \rm{MW/cm^2}$, it undergoes a saturation and eventually decreases at higher
intensities. For clarity purposes, in Fig.5(b) we plot the efficiency $\eta_t$ as a function of $I^{(1)}_i$ for $\theta = 40.9^\circ$. Such a
behavior of $\eta_t$ is evidently a consequence of the fact that, at low optical intensity, the undepleted pump regime rules the phenomenology and
provides the growing efficiency $\eta_t$ whereas, at higher intensities, the full nonlinear regime comes into play where the FW and the SHW
experience their mutual nonlinear interaction. It is however remarkable that, in the situation considered in Fig.5, the maximum efficiency
$\eta_t=0.316$ is achieved at the intensity $I^{(1)}_i \simeq 65 \: \rm{MW/cm^2}$ (see Fig.5(b)) which is considered to be a small value in a
nonlinear optical setup.
\begin{figure}
\includegraphics[width=0.45\textwidth]{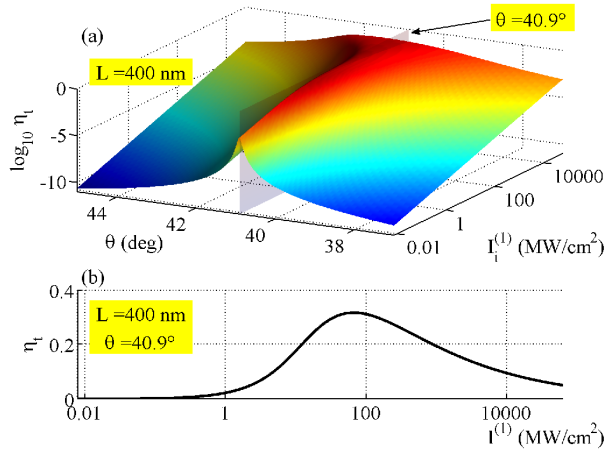}
\caption{(Color online) Efficiency $\eta_t$ of a $400 \: \rm{nm}$ thick slab of converting the incident FW into the transmitted SHW. (a) Logarithmic
plot of $\eta_t$ as a function of the incidence angle $\theta$ and the optical intensity $I^{(1)}_i$ of the incident FW. (b) Linear plot of $\eta_t$
at $\theta = 40.9^\circ$ (detailed from the shadowed section of panel (a)). }
\end{figure}

In order to prove the feasibility of the above described SHG process, we discuss here the numerical design of a nano-structured composite
characterized by the above unconventional features. The slab (see. Fig.1) is manufactured by alternating, along the $y-$axis, two kind of layers
(labelled with $a$ and $b$) of thicknesses $d_a=11.84 \: \rm{nm}$ and $d_b = 88.16 \: \rm{nm}$, respectively. We choose the layers of kind $a$ to be
filled with alpha-iodic acid ($\alpha$-$\rm{HIO}_3$) crystals with principal axis along the cartesian axis so that the permittivities relevant for
our purposes are $\epsilon_{ax}^{(1)} = 3.7588$, $\epsilon_{az}^{(1)} = 3.8517$, $\epsilon_{ay}^{(2)} = 3.5871$ and the non-vanishing components of
the second-order nonlinear optical tensor are $d_{14} = d_{25} = d_{36} = \chi_a /2 = 7 \: \rm{pm/V}$ \cite{Dmitri}. For the layers of kind $b$ we
choose a polymer (e.g. poly(methyl methacrylate), PMMA) with permittivities $\epsilon_h^{(1)} = 2.2022$ and  $\epsilon_h^{(2)}= 2.2575$  hosting
silver-coated spherical PbSe quantum dots (QDs) with inner radius $R_1=5 \: \rm{nm}$ and outer radius $R_2=6.38 \: \rm{nm}$. Exploiting the
electrostatic approach of Ref.\cite{ZengZe}, the structured particle can be described by the equivalent permittivity $\epsilon_S = \epsilon_{Ag}
[\epsilon_{QD} (1+2\rho) +2 \epsilon_{Ag} (1-\rho)] /  [\epsilon_{QD} (1-\rho) +2 \epsilon_{Ag} (2+\rho)]$ where $\rho = R_1^3 / R_2^3$,
$\epsilon_{Ag} = \epsilon_\infty - \omega_p^2 /(\omega^2+i \Gamma \omega)$ is the Drude-type silver permittivity ($\epsilon_\infty = 4.56$, $\omega_p
= 1.38 \cdot 10^{16} \: \rm{s^{-1}}$ and $\Gamma = 0.1 \cdot 10^{15} \: \rm{s^{-1}}$ \cite{Scatte}) and $\epsilon_{QD} = \epsilon_b + A\omega_0^2 /
(\omega^2 - \omega_0^2 + i 2\gamma \omega)$ is the dielectric permittivity of the considered semiconductor QD ($\epsilon_b=12.8$, $\omega_0 = 2.2793
\cdot 10^{15} \: \rm{s^{-1}}$, $\gamma = 1.2788 \cdot 10^{12} \: \rm{s^{-1}}$ \cite{FuFuFu} and $A=9.6 \cdot 10^{-4}$). Assuming $F_{QD}=0.1$ for the
volume filling fraction of the QSs embedded in the polymer and using the standard Maxwell Garnett mixing rule for obtaining the effective dielectric
permittivity, we obtain, for the layers of kind $b$, $\epsilon_{b}^{(1)} = -0.5160 + 0.0002i$ and $\epsilon_{b}^{(2)} = 2.7819 + 0.0127i$. Since the
period $d_a + d_b = 100 \: nm$ is considerably smaller than the two wavelengths, FW and SHW propagating through the slab will experience the effect
of a homogeneous medium characterized by the permittivities $\epsilon_x^{(1)}= f_a \epsilon_{ax}^{(1)} + f_b \epsilon_{b}^{(1)}= -0.0099+0.0001i$,
$\epsilon_z^{(1)}= f_a \epsilon_{az}^{(1)} + f_b \epsilon_{b}^{(1)}= 0.0011+0.0001i$, $\epsilon_y^{(2)}= [f_a / \epsilon_{ay}^{(2)} + f_b /
\epsilon_{b}^{(2)}]^{-1}= 2.8579+0.01184i$ (where $f_a= d_a/(d_a+d_b)$ and $f_b = d_b/(d_a+d_b)$ are the layers volume filling fractions), whereas
the effective susceptibility is $\chi = f_a (\epsilon_y^{(2)}/\epsilon_{ay}^{(2)}) \chi_a = 1.3206+0.0054i \: \rm{pm/V}$ (see Ref.\cite{BoydBo} for
the derivation of these relations). Even though theoretical (note that we have neglected the uncertainties of all the used quantities), the proposed
approximate design reveals that, by suitably acting on free parameters (kind of media, QDs volume filling fractions, inversion population factor A,
etc.), a medium for observing the proposed efficient SHG can actually be conceived.

In conclusion, we have identified a novel mechanism able to provide a strong field enhancement in anisotropic optical metamaterials having very small
cartesian permittivities of different signs. We have showed that, if the slab has a nonlinear quadratic response, the field enhancement is
responsible for an efficient SHG process even if the slab thickness is smaller than the wavelength of the SHW. It is worth stressing that the
proposed field enhancement mechanism can easily be observed also in setups different from the considered slab. Our findings offers an hitherto
unexplored way for achieving efficient SHG from nanostructures and therefore we believe they could pave the way for conceiving a novel generations of
nanometric-sized coherent light sources.



\begin{thebibliography} {aa}
\bibitem{Higher} We are neglecting here the genereration of higher harmonics.
\end{thebibliography}
\end{document}